%% file: main.tex
\documentclass[10pt, conference]{IEEEtran}  
\usepackage{enumitem}
\usepackage{makecell}
\usepackage{multirow}
\usepackage{xspace}
\usepackage[linesnumbered,ruled,vlined]{algorithm2e}
\usepackage{booktabs}
\usepackage{pifont}
\usepackage{fancyvrb}
\usepackage[hidelinks]{hyperref}
\usepackage{subcaption}

\usepackage{enumitem}
\setlist[enumerate]{leftmargin=1.5em}
\setlist[itemize]{leftmargin=1.5em}

\newtheorem{theorem}{Theorem}
\newtheorem{definition}[theorem]{Definition}

\newcommand{\name}{Diff\-Test\-Gen}
\newcommand{\code}[1]{\texttt{\small#1}}
\newcommand{\scode}[1]{\texttt{\footnotesize#1}}
\newenvironment{codeblock}{\Verbatim[fontsize=\footnotesize, frame=none, commandchars=\\\{\}]}{\endVerbatim}

\input{tex_files/macros.tex}

\usepackage[most]{tcolorbox}
\newtcolorbox{cbox}{%
    enhanced,
    colback=green!10,
    colframe=green!30!black,
    arc=1mm,
    boxrule=0.6pt,
    left=2mm, right=2mm, top=2mm, bottom=2mm
}

\newtcolorbox{promptMSG}[3][]{%
  enhanced, attach boxed title to top right={yshift=-\tcboxedtitleheight}, 
  boxed title style={ size=small, 
    colback=gray!20, colframe=gray!20, sharp corners=downhill, 
    arc=.4cm, top=1mm,bottom=1mm,left=1mm,right=1mm}, 
  fonttitle=\color{black}\itshape, 
  colframe=#3, colback=#3, 
  top=\tcboxedtitleheight, bottom=0mm, sharp corners=downhill, 
  arc=.4cm, title={#2},#1 
}

\newtcolorbox{promptMSGNoTitle}[2][]{%
  enhanced, 
  colframe=#2, colback=#2, 
  top=1mm, bottom=1mm, left=1mm, right=1mm, 
  sharp corners=downhill, 
  arc=.4cm, 
  #1 
}

\def\BibTeX{{\rm B\kern-.05em{\sc i\kern-.025em b}\kern-.08em
    T\kern-.1667em\lower.7ex\hbox{E}\kern-.125emX}}

\definecolor{darkorchid}{rgb}{0.6, 0.2, 0.8}

\begin{document}
\title{DiffTestGen: Change-Directed LLM-Based Testing for Exposing Behavioral Differences}

\author{\IEEEauthorblockN{1\textsuperscript{st} Huimin Hu}
\IEEEauthorblockA{\textit{CISPA Helmholtz Center} \\
\textit{ for Information Security}\\
Stuttgart, Germany \\
huhuimin236@gmail.com}
\and
\IEEEauthorblockN{2\textsuperscript{nd} Cristian Cadar}
\IEEEauthorblockA{\textit{Department of Computing} \\
\textit{Imperial College London}\\
London, United Kingdom \\
c.cadar@imperial.ac.uk}
\and
\IEEEauthorblockN{3\textsuperscript{rd} Michael Pradel}
\IEEEauthorblockA{\textit{CISPA Helmholtz Center} \\
\textit{ for Information Security}\\
Stuttgart, Germany \\
michael@binaervarianz.de}
}

\maketitle

\begin{abstract}
Software testing plays a critical role in maintaining software quality.
As software evolves over time, it is important to ensure that any behavioral changes occur as intended by developers. 
A promising approach for this goal is to generate tests that expose behavioral differences between the old and new versions of a program.
However, current approaches fail to trigger behavioral differences for many code changes.
This paper presents~\name{}, a novel change-directed, LLM-based differential testing approach specifically designed to expose behavioral differences introduced by a code change.
The approach is enabled by two key contributions:
First, \name{} leverages static call graph analysis and project documentation to identify valid entry points for test generation and to guide the LLM toward reaching the changed code.
Second, \name{} iteratively improves our newly introduced union coverage metric, which combines coverage of modified code in the old and the new version, by providing targeted coverage feedback to the LLM.
We evaluate \name{} on two datasets comprising a total of \totalsize{} PRs.
\name{} exposes behavioral differences in \totalrate{} of the PRs while achieving an average union coverage of \totalunioncov{}. 
Compared with the baselines, \name{} exposes \totaldiffermore{} more PRs overall and increases code coverage by \chacocovmore{} and \testoracovmore{} percentage points, respectively. 
By integrating \name{} with the Testora regression detector, we show that the identified behavioral differences can be used to detect regression bugs missed by the best existing approaches.
\end{abstract}


\section{Introduction}
\label{sec:intro}
Software systems continuously evolve through code changes that implement new features, fix defects, improve performance, or refactor existing implementations. 
While developers often intend these changes to affect specific aspects of a program, code changes can also introduce behavioral changes that extend beyond their intended scope. 
Such changes may alter the observable behavior of a system under particular inputs or execution conditions, potentially impacting correctness, compatibility, reliability, or user experience. 
Understanding the behavioral consequences of code changes is therefore a fundamental challenge in software evolution and maintenance.

Several techniques have been proposed to support software evolution, including automated test generation~\cite{FSE24_ChatUniTest, FSE25_CoverUp, ICSE25_IntentionGuided, 2026AdvancingCodeCoverage}, regression testing~\cite{FSE24_Code_Aware_Prompting,le2026testweaver}, and differential testing~\cite{danglot2020approach, ASE23_DifferentialPrompting, SANER24_TraceJIT}. 
Automated test generation synthesizes tests to exercise source code, program behavior, or requirements.
Regression testing re-executes existing tests to ensure that modifications do not introduce regressions.
Differential testing identifies differences by comparing the behavior of multiple program versions.

If effective, differential testing provides a promising foundation for automatically exposing behavioral differences between program versions.
Unlike traditional testing approaches that assess correctness against a specification, differential testing focuses on identifying discrepancies by executing multiple program variants under identical inputs and comparing their outputs or behavior. 
Applied to software evolution, it can systematically reveal inputs that trigger differences between program versions before and after code changes. 
For example, the recent Testora approach~\cite{pradel2025testora} uses differential testing to detect regression bugs by identifying behavioral changes that contradict the developer's intent documentation in the description of a pull request (PR).
However, such regression testing approaches rely on the availability of tests that exercise the changed code, which is often not the case in practice.

Despite significant advances in the proposed techniques, systematically exposing behavioral differences introduced by code changes remains challenging. 
First, many existing test generation techniques are not explicitly guided by code changes~\cite{FSE25_CoverUp, ICSE25_IntentionGuided, FSE2026_TestGeneralizer}, 
and therefore may fail to exercise newly modified program behavior. 
Second, changed code regions are often difficult to reach, e.g., because the modified function is not directly reachable from a public entry point into the project.
Together, these challenges limit the ability of existing approaches to systematically expose behavioral differences introduced by code changes, which in turn limits the ability of downstream techniques~\cite{pradel2025testora} to detect regression bugs.

\begin{figure}[t]
  \centering
  \hspace{0.12em}\includegraphics[width=0.485\textwidth]{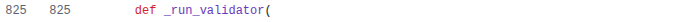}
  \includegraphics[width=0.49\textwidth]{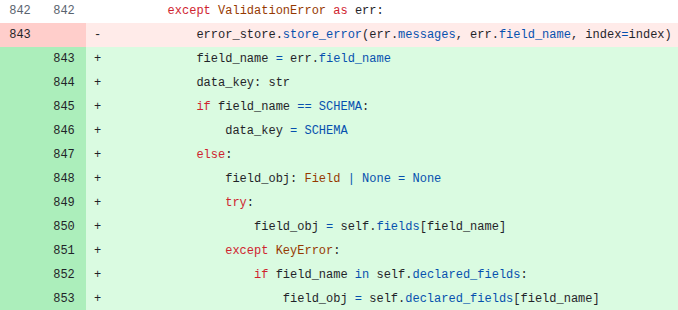}
  \caption{Motivating example.}
  \label{fig:motivating_example}
\end{figure} 

Figure~\ref{fig:motivating_example} illustrates a motivating example of a code change in a private function \code{\_run\_validator}.
Although the change is localized, exposing its behavioral impact is far from straightforward. 
Since the private function is not directly accessible, exercising the modified code requires a detailed understanding of how to reach the changed private function. 
While LLMs have shown promising results for generating tests, they often lack sufficient contextual information to determine how the changed code can be exercised effectively.
Consequently, the changed code may remain unexplored, and the behavioral differences introduced by the modification may not be observed.

To address limitations of prior work in effectively performing differential testing of code changes, we present \name{}, a change-directed, LLM-based testing approach for exposing behavioral differences between two program versions. 
Given a PR, \name{} generates and executes tests to expose behavioral differences between the original and changed program versions.
Specifically, it uses static call graph analysis to identify accessible entry points that can be used to reach changed functions. 
This helps guide test generation toward triggering changed code lines.
In addition, it leverages reference tests and their coverage information as feedback to guide the exploration of previously unexecuted changed lines.
By making behavioral differences explicit, \name{} could support a wide range of development activities, including regression investigation, software maintenance, and code review.

To evaluate our work, we apply \name{} to code changes made in a total of 463 PRs from popular open-source projects. 
\name{} exposes behavioral differences in \totalrate{} of the PRs, while achieving an average union coverage of \totalunioncov{}.
The analyzed PRs are from two datasets from prior work~\cite{pradel2025testora,ICSE026_chaco}, enabling us to directly compare \name{} with the best existing approaches.
On the two datasets, our approach detects behavioral differences in \chacodifferoutputrate{} and \testoradifferoutputrate{} of the PRs, with an average union coverages of \chacounioncov{} and \testoraunioncov{} on the datasets, respectively. 
Compared with the baselines, \name{} exposes \totaldiffermore{} more PRs overall and increases code coverage by \chacocovmore{} and \testoracovmore{} percentage points, respectively. 
Integrating \name{} with the Testora regression detector~\cite{pradel2025testora}, we further show that the identified behavioral differences can be used to detect regression bugs missed by the best existing approaches.


In summary, this paper makes the following contributions:
\begin{itemize}[leftmargin=*]
\item We study the problem of generating behavior-exposing tests for code changes, identifying two key challenges that limit the effectiveness of existing approaches.
\item We present a change-directed LLM-based testing approach that leverages information related to code changes, public API documentation, and static call graph analysis to guide LLMs in producing tests capable of exposing behavioral differences between two versions of a program. 
\item We conduct a large-scale experimental evaluation demonstrating the effectiveness of the proposed approach in exposing behavioral differences.
\item To foster future work, we make all code and data associated with \name{} publicly available: \url{https://github.com/sola-st/DiffTestGen}
\end{itemize}

\section{Approach}
\label{sec:approach}

\subsection{Problem Definition}
\label{subsec:prob_definition}

We address the problem of generating tests that expose behavioral differences between two program versions, where one version is obtained by applying a code change to the other.

\begin{definition}[Change-directed testing]
Given a PR and its associated code base, change-directed testing creates a set~$T$ of test cases that expose behavioral differences between the pre-change and post-change versions of the code.
\end{definition}

The above definition relies on the following notion of behavioral difference between two program versions:

\begin{definition}[Behavioral difference]
\label{def:behavioral_difference}
Given a test case~$t$, let $o_{old}$ and $o_{new}$ be the outputs produced by executing $t$ on the pre-change and post-change versions of a code base, respectively, and let $e_{old}$ and $e_{new}$ be the type of runtime errors (if any) produced by executing $t$ on the pre-change and post-change versions, respectively.
A behavioral difference exists if one of the following conditions holds:
\begin{itemize}
    \item Both executions produce runtime errors, but $e_{old} \neq e_{new}$.
    \item Only one execution produces a runtime error.
    \item Neither execution produces an error, but $o_{old} \neq o_{new}$.
\end{itemize}
\end{definition}

When comparing test behavior, we ignore incidental and flaky differences that do not reflect meaningful behavioral differences.
Specifically, we normalize the outputs by removing incidental information, such as timestamps.
Moreover, we re-execute each generated test on both the old and new program versions. 
If the execution result differs in either version from that observed in the initial run, the test is considered flaky and is discarded.
Otherwise, the test is regarded as stable, and any observed difference between the two program versions is reported as a behavioral difference.

\emph{Scope of considered code changes.}
\name{} targets code changes made in PRs of Python projects.
Specifically, we consider PRs that satisfy all of the following criteria:
\begin{itemize}[leftmargin=*]
  \item To focus on changes that are more likely to affect program behavior, we consider a PR only if it modifies at least one non-test Python source file.
  \item To focus on behavioral differences, we exclude changes that modify only comments or documentation.
  \item To mitigate potential scalability issues, we restrict modifications to at most three non-test source files.
  \item To isolate changes introduced by the PR, we require that it corresponds to a commit with exactly one parent (i.e., it is not a merge commit).
\end{itemize}

\subsection{Overview}
\label{sec:overview}

\begin{figure*}[t]
  \centering
  \includegraphics[width=0.94\textwidth]{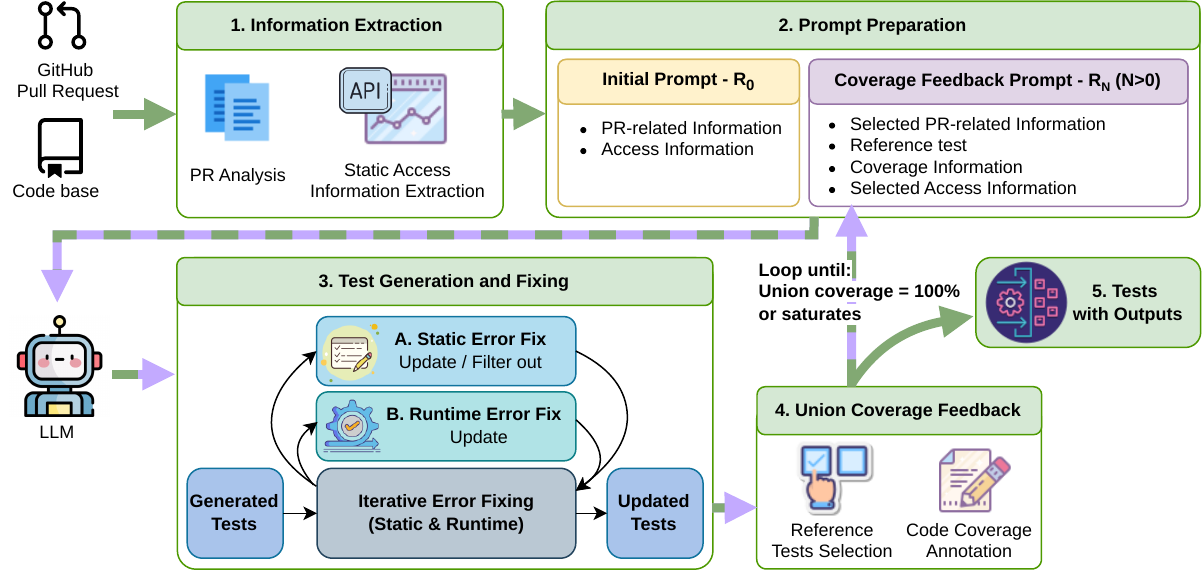}
  \caption{Overview of \name{}.} 
  \label{fig:test_gen_overview}
\end{figure*}


Figure \ref{fig:test_gen_overview} summarizes our change-directed LLM-based testing approach.
The inputs to \name{} are a GitHub PR and its code base.
The process begins by extracting PR-related information~(\textbf{1}), including analyzing the PR and collecting static access information. 
Next, it constructs an initial prompt~(\textbf{2}) and sends it to an LLM to generate tests.
The generated tests are then refined through an iterative error-fixing phase~(\textbf{3}), consisting of two inner loops: 
one that resolves static validity errors in the generated tests~(\textbf{3A}), and another that addresses runtime errors identified during test execution~(\textbf{3B}). 
\name{} also includes an outer coverage feedback loop (with connections between phases indicated by bicolored arrows) that keeps track of how much of the changed code has been covered. 
Using the updated tests and execution results from Phase 3, it checks the achieved coverage~(\textbf{4}) to determine whether to initiate another round of test generation. 
If the coverage reaches 100\% or remains unchanged compared with the prior round, \name{} outputs the generated tests and stops (\textbf{5}). 
Otherwise, it selects a reference test, annotates the code with coverage information, and prepares a coverage feedback prompt to resume another round of test generation. 

In the rest of this section, we describe analysis of code change and static access information extraction (\S\ref{subsec:approach_change_access_info}), construction of test generation prompt (\S\ref{subsec:approach_prompt}), 
static validity check and runtime error feedback loop (\S\ref{subsec:approach_inner_feedback}), and finally outer coverage feedback loop for subsequent rounds (\S\ref{subsec:approach_refer_test}).

\subsection{Analysis of Code Change and Access Information}
\label{subsec:approach_change_access_info}
To provide the model with information about code changes that helps generate tests that expose behavioral differences, \name{} first identifies what changed and then determines how the changed code can be reached.

\subsubsection{Change Analysis}
\label{subsubsec:approach_change_analysis}

We start by analyzing the change.
To obtain the \emph{diff} of a PR, we query the GitHub API. 
Based on the diff, \name{} performs an AST-based analysis to extract the changed functions, including their names and bodies. 

Next, to address the challenge of reaching the changed code, we first categorize the changed functions, which enables subsequent stages to extract information on how to invoke each changed function. 
In detail, we categorize functions\footnote{In this paper, we use \emph{function} for general procedures, and \emph{method} for functions defined within classes, following language-specific terminology.} into three types based on their accessibility:
(i)~changed public functions $\mathit{f}_{pub}$, (ii) changed private functions $\mathit{f}_{pri}$, or (iii)~changed special methods $\mathit{f}_{spe}$, which are intended to be invoked only through their corresponding operations. 
To determine the category of a changed function, we consider both naming conventions and the list of public APIs specified by the projects. 
The public API approach leverages the fact that large projects generally maintain well-structured documentation, which includes an explicit list of their public APIs. 
For modules documented in the public API list, a function is considered public if it appears in the list, and private if it does not. 
For modules not covered in the public API documentation, naming conventions are applied: 
a function is classified as private if its name begins with a single underscore ``\_'' but not a double underscore ``\_\_'', and public otherwise.
A function is classified as a special method if its name begins with ``\_\_''.

\subsubsection{Access Information}
\label{subsubsec:approach_access_information}



\input{tex_files/changed_function_categorization.tex}

Providing the model with information about how to reach the changed functions, called \emph{access information}, is crucial for generating tests that can exercise the modified code.
The exact access information we provide depends on the category of the changed function; Table~\ref{tab:changed_function_types} summarizes the details for changed public functions, changed private functions, and changed special methods. 
One common component is \emph{import lines}, i.e.\ lines of the form \code{from ...}\code{import ...} that indicate how the target symbol should be loaded.
For example, an import line such as \code{from scipy.sparse import random\_array} specifies explicitly how to load the function \code{random\_array}. 

\input{tex_files/access_info_examples.tex}
Figure~\ref{fig:access_info_examples} provides illustrative examples of the access information associated with each category.
Figure~\ref{fig:access_info_example_public} shows an example for changed public functions, where \code{style} is identified as public based on the public API documentation check and belongs to class \code{DataFrame}. Its access information includes the relevant import line, together with the signature and docstrings of class \code{DataFrame}.
Figure~\ref{fig:access_info_example_private} presents an example for changed private functions, which corresponds to the case shown in Figure~\ref{fig:motivating_example}; in this case, the access information is defined with respect to a publicly accessible entry function $\mathit{f}_\mathit{entry}$ from which the changed private function can be accessed, so that generated tests exercise only the project's public (user-accessible) APIs.
For example, the changed private function is \code{\_run\_validator}, and one such entry function is \code{validate}. We therefore provide the signatures and docstrings of these entry functions, and if an entry function belongs to a class (e.g., the class \code{Schema}), we additionally include the class's signature and docstring.
This access information enables \name{} to address the second challenge (changed code regions are often hard to reach) mentioned in \S\ref{sec:intro}.
Figure~\ref{fig:access_info_example_special} presents an example for changed special methods in which the special method is \code{\_\_call\_\_}. For such methods, the access information is defined with respect to the base class in which the changed special method is defined, including the import line, the signature and docstrings of the base class, as well as the usage guideline (``hints'') on how the special method can be invoked.
\subsubsection{Extracting Access Information}
\label{subsubsec:approach_extracting_access_information}

To extract the access information for changed functions, we perform static analysis on the modified code, focusing on identifying import lines, obtaining the signatures and docstrings of both functions and classes, and in the case of private functions also extracting call paths. 
We first construct a static call graph, defined as a directed graph in which nodes represent functions and edges represent call relationships between functions. 
Based on this graph, we derive call paths that describe how one function can reach another. 
These call paths contribute to a key component of the access information used to guide the model toward the changed functions. 

\input{tex_files/call_path_algorithm.tex}

For a changed private function, Algorithm~\ref{alg:access_information} traces backward from the changed private function, following its callers until reaching publicly accessible entry functions, and returns a list of access information, each associated with a publicly accessible entry function. 
It begins by initializing a queue with the changed function (line~\ref{alg:init_queue}) and then repeatedly expands callers in the \code{while} loop (lines~\ref{alg:while_start}--\ref{alg:while_ends}). 
The exploration of a call path stops once a publicly accessible entry function is reached (lines~\ref{alg:ends_check_starts}--\ref{alg:ends_check_ends}), while recursive cycles are filtered out to avoid revisiting the same path (lines~\ref{alg:loop_handle_starts}--\ref{alg:loop_handle_ends}). 
After this search phase, Step 2 groups the discovered paths by entry function and keeps the shortest one per group (lines~\ref{alg:2.1_starts}--\ref{alg:2.1_ends}), then selects the top-$k$ shortest remaining paths (line~\ref{alg:2.2}, with $k$=$5$), yielding a compact set of access information that captures the most direct ways to reach the changed private function.

For a changed public function, this static analysis directly yields the required access information. 
For a changed special method, it identifies the base class in which the changed special method is defined and yields the required access information, while the hint information is obtained from the official Python documentation.

\subsection{Initial Prompt Construction}
\label{subsec:approach_prompt}
We construct two different kinds of prompts: a prompt for the initial test generation round ($R_0$) and a coverage feedback prompt for subsequent coverage feedback rounds ($R_n, n>0$). 
Table~\ref{tab:prompt_info} presents the information contained in the prompts, ordered according to its appearance within each prompt. 
In this section, we focus on the prompt for the initial test generation round ($R_0$).

As described in \S\ref{subsec:approach_change_access_info}, we obtain the diff and extract the changed function and corresponding access information.
By providing the diff and explicitly instructing the model to ``expose behavioral differences introduced by the diff'', we guide test generation toward change-related behaviors, helping address the first challenge described in \S\ref{sec:intro}. 
We categorize changed functions into test and non-test functions to focus on actual changes to the SUT, which are primarily reflected in non-test functions. 
Specifically, leveraging the naming conventions of Python test files (i.e., filenames prefixed with ``test\_'' or suffixed with ``\_test''), 
we classify functions defined in non-test Python files as non-test functions, and those defined in test files as test functions. 
We also provide the changed test functions in the prompt, to show the models how to access the changed non-test functions. 

\input{tex_files/prompt_info.tex}

As shown in Table~\ref{tab:prompt_info}, for $R_0$, the required information items provide the project name, the diff, and the fully qualified names of the changed functions, while the bodies of non-test functions, access information, and the bodies of test functions are added when they fit within the target prompt length.

\subsection{Inner Feedback Loops}
\label{subsec:approach_inner_feedback}
After the model generates tests, \name{} refines them through two inner feedback loops: static validity check and runtime error feedback.

\subsubsection{Static Validity Check}
\label{subsec:approach_static_feedback}
To improve the executability and quality of the generated tests, \name{} first performs a static validity check.
For each generated test, \name{} verifies that the code can be parsed into an AST using Python's \code{ast} module and determines whether it invokes any private functions.
If a test contains static validity errors, we send both the code and the corresponding error message back to the model for correction.
A static validity error is considered resolved once the test becomes parsable and no longer calls private functions.
We allow up to five attempts of static validity checking. After that, we retain only the tests that have been successfully corrected and discard any that still contain errors.

\subsubsection{Runtime Error Feedback}
\label{subsec:approach_runtime_feedback}
To examine the tests, we identify two commits associated with a PR:
the commit before the PR is created ($\mathit{c_{old}}$), and the commit where the PR is applied ($\mathit{c_{new}}$).
We then create an isolated execution environment for each commit, marked as $\mathit{env_{old}}$ and $\mathit{env_{new}}$, respectively.
For all tests passing the static validity check, we perform a runtime check by executing them in these environments and recording any runtime errors.
If a test fails, we return the test code and the corresponding error message to the model for correction.
As with the static validity checks, we allow up to five attempts to resolve each runtime error.
A test is considered to have passed the runtime check if it executes successfully on at least one of the two commits.
Unlike the static validity check step, we keep all executions, even those that still fail after five attempts, because failing test executions can still provide useful information about code behavior and coverage.

\subsection{Outer Coverage Feedback Loop}
\label{subsec:approach_refer_test}
\label{subsec:union_coverage_section}

After the inner feedback loops, \name{} uses coverage information to decide whether to initiate another test generation round.
We compute a line-level coverage of changed Python code in the PR for each test execution,
and define the union coverage of a single test as follows: 


\begin{definition}[Union Coverage]
\label{def:union_coverage_def}
Given a PR, let $\mathit{Num}_{\text{changed}}^{\text{old}}$ and $\mathit{Num}_{\text{changed}}^{\text{new}}$ denote the number of changed Python code lines in the pre-change and post-change program versions, respectively. 
For the executions, let $\mathit{Num}_{\text{covered}}^{\text{old}}$ and $\mathit{Num}_{\text{covered}}^{\text{new}}$ denote the number of changed Python code lines covered in the pre-change and post-change versions, respectively.
The union coverage of a single test is:
\setlength{\abovedisplayskip}{2.5pt}
\setlength{\belowdisplayskip}{2.5pt}
\[
\mathit{Cov}_{\text{union}}^{\text{test}} =
\frac{
\mathit{Num}_{\text{covered}}^{\text{old}} + \mathit{Num}_{\text{covered}}^{\text{new}}
}{
\mathit{Num}_{\text{changed}}^{\text{old}} + \mathit{Num}_{\text{changed}}^{\text{new}}
}
\]

\end{definition}

We consider a changed Python code line in the computation only if it is located in a non-test Python file and is executable (i.e., not comments or docstrings). 

\input{tex_files/refer_test_example.tex}

To guide the model to generate tests that can cover more changed code lines, \name{} selects a test from the prior round as a \emph{reference test} for the current round.  
Intuitively, \name{} selects the prior test that already covered lines closest to the remaining uncovered changed lines, because that test is likely to need the smallest adjustment to reach the uncovered code.
Figure~\ref{fig:reference_test} shows an example of a reference test and its coverage information. 
Motivated by the idea of using a natural way for the model to understand coverage information,
\name{} encodes the coverage information as comments (e.g., \code{\# COVERED} and \code{\# TO\_COVER}) within the affected functions.  


After selecting a reference test and annotating the corresponding non-test functions with coverage information, \name{} constructs a coverage feedback prompt for the next round of test generation. 
As shown in Table~\ref{tab:prompt_info}, each coverage feedback prompt ($R_n$, $n>0$) includes the project name, the diff, and the fully qualified names of the changed functions, as in the initial prompt ($R_0$). 
The coverage feedback prompt further restricts the access information and non-test function bodies to those associated with uncovered changed functions, while also applying the target prompt length constraint.

\section{Evaluation}

\subsection{Research Questions}
\label{section:RQ}
In our evaluation, we investigate four research questions:
\begin{enumerate}[label=RQ\arabic*, leftmargin=2.4em]
  \item \textbf{Effectiveness:} How effectively does \name{} expose behavioral differences resulting from code changes, compared with baseline approaches?
  \item \textbf{Component Contributions:} What is the contribution of each component of \name{} to overall effectiveness?
  \item \textbf{Efficiency:} How efficiently does \name{} generate differentiating tests in terms of execution time, token consumption, and overall financial cost?
  \item \textbf{Usefulness:} How useful is \name{} in detecting regression bugs?
\end{enumerate}

\subsection{Experimental Setup}
\subsubsection{Baselines}
\label{subsec:baselines}
We compare \name{} against three baselines: Testora~\cite{pradel2025testora}, Testora++, and ChaCo~\cite{ICSE026_chaco}. 
Testora is an automated technique for detecting unintended behavioral changes by leveraging natural language information associated with a code change as a test oracle. 
We select Testora as a baseline because its test generation phase targets identifying differences between code versions, which aligns with \name{}.
By default, both Testora and \name{} generate 20 tests in the initial round. 
However, \name{} generates additional tests in its subsequent rounds, whereas Testora does not, resulting in more tests generated by \name{}.
To control for potential bias introduced by this difference, we introduce Testora++, which increases the number of generated tests in the initial round of Testora from 20 to 100. 

ChaCo is a pull request-based test augmentation technique that generates tests to increase the coverage of an existing test suite and integrate the generated tests into that suite. 
\name{} has a different primary objective, as it focuses on exposing the behavioral differences introduced by code changes.
For a fair comparison, we consider the coverage contributed by ChaCo-generated tests and measure whether these tests reveal differences in outputs between the old and new code versions. 

\subsubsection{Dataset}
We evaluate \name{} using two datasets from prior work Testora and ChaCo, which together cover four open-source Python projects.
Table~\ref{tab:datasets} summarizes the datasets, including the number of PRs and the average size of the PRs in terms of changed non-test Python code lines. 
The total number of PRs is \totalsize{}, as the two datasets overlap by 10 PRs that appear in both.

\begin{table}[t]
  \centering
  \caption{Summary of datasets}
  \begin{tabular}{@{}lrrrr@{}}
    \toprule
    Project & \multicolumn{2}{c}{Testora data} & \multicolumn{2}{c}{ChaCo data} \\ 
    \cmidrule(lr){2-3} \cmidrule(lr){4-5}
    & PRs & Avg. change size & PRs & Avg. change size \\
    \midrule
    keras  & 133 & 13.3 & -- & -- \\
    marshmallow   & 53  & 10.4 & -- & --\\
    pandas  & 127 & 6.6 & 16 & 7.6 \\
    scipy  & 126 & 16.0 & 18 & 48.5\\
    \midrule
    Total & \testoracheckedsize{} & 11.8 & \chacocheckedsize{} & 29.3 \\
    \bottomrule
  \end{tabular}
  \label{tab:datasets}
\end{table}

\paragraph{Testora data}
We use the dataset originally constructed in Testora~\cite{pradel2025testora} for our evaluation. 
In addition to the filtering criteria of Testora, \name{} further excludes PRs that involve only non-Python code changes, as it targets Python-related changes (see \S\ref{subsec:prob_definition}).
This results in a final dataset of \testoracheckedsize{} PRs, with an average of 11.8 changed non-test Python code lines per PR.

\paragraph{ChaCo data}

To keep consistency and focus on our target projects, we use the \emph{pandas} and \emph{scipy} PRs from the ChaCo dataset~\cite{ICSE026_chaco}. 
Specifically, we retain only the PRs for which ChaCo successfully integrated at least one generated test into the corresponding test files. 
The integrated tests are required to execute successfully without runtime errors or test failures, and to achieve a non-zero increase in code coverage.
Furthermore, we exclude tests that invoke private functions. 
This yields a final evaluation dataset of \chacocheckedsize{} PRs, with an average of 29.3 changed non-test Python code lines per PR.


\subsubsection{Evaluation Metrics}
To evaluate \name{}'s ability to expose the behavioral differences of the given PRs, we use the following two metrics: 
\paragraph{Number of PRs with behavioral differences ($\mathit{Num}_\text{PR}$)}
First, we measure the number of PRs with exposed behavioral differences. 
A single PR may have multiple generated tests that expose behavioral differences, and multiple tests may reveal the same behavioral difference.
To avoid redundant counting, we use $\mathit{Num}_\text{PR}$ as a primary metric. 
For selected analyses, we also report $\mathit{Num}_\text{tests}$, representing the number of tests that expose behavioral differences, as a supplementary metric.

\paragraph{Union coverage ($\mathit{Cov}_\text{union}$)} 
Second, as introduced in \S\ref{subsec:union_coverage_section}, we report the overall union coverage across all PRs,  where N is the total number of PRs in the dataset:
{
\setlength{\abovedisplayskip}{2.5pt}
\setlength{\belowdisplayskip}{2.5pt}
\[
\mathit{Cov}_\text{union} = \frac{1}{N} \sum_{i=1}^{N} \mathit{Cov}_\text{union}^{\text{test}^i}
\]
}




\subsubsection{Ablation Study}
To obtain a comprehensive understanding of how \name{} exposes behavioral differences and how each component contributes to its effectiveness, we evaluate several variants of \name{}.
We implement \name{} on top of Testora~\cite{pradel2025testora}, adding around 2,800 lines of code in its core components. 
Each variant incrementally extends Testora by introducing an additional feature, allowing us to isolate and assess the contribution of individual components.
The variants are configured as follows:
\begin{itemize}
\item \textit{\onlycg{}.} This variant extends Testora by incorporating access information (described in~\S\ref{subsec:approach_change_access_info}).
\item \textit{\onlycov{}.} This variant extends Testora by adding a coverage feedback loop with a reference test and its union coverage information (described in ~\S\ref{subsec:approach_refer_test}).
\end{itemize}

\subsubsection{Large Language Models}
\label{subsec:llm}
We use \emph{gpt-5-mini} as the default model in the experiments. The baselines originally use \emph{gpt-4o-mini}.
Testora~\cite{pradel2025testora} is directly comparable to \name{}, and we therefore re-run Testora using \emph{gpt-5-mini} to ensure a fair comparison under the same model configuration.
Because we were unable to re-run ChaCo~\cite{ICSE026_chaco}, we additionally evaluate \name{} with \emph{gpt-4o-mini} to ensure consistency in this comparison.

\begin{figure*}[t]
    \centering

    \begin{subfigure}{\textwidth}
        \centering
        \includegraphics[width=\textwidth]{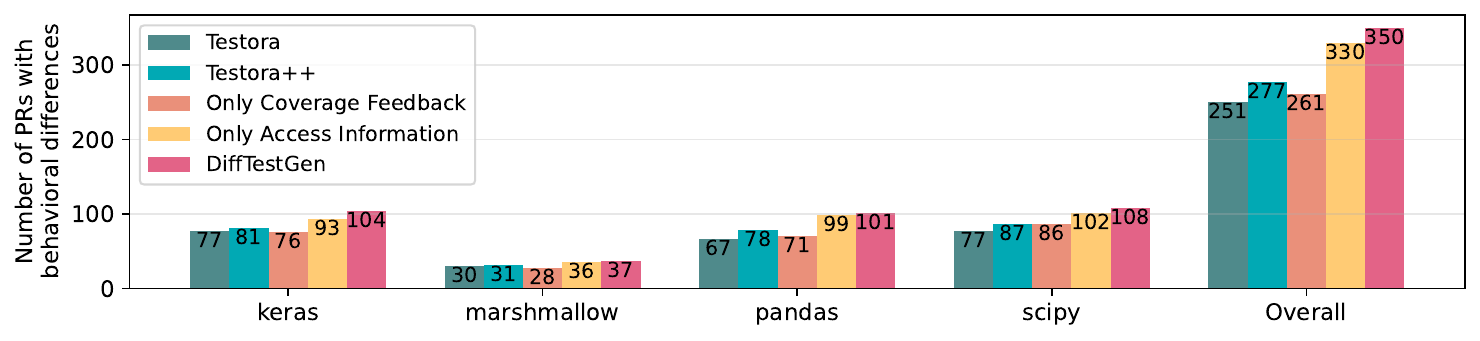}
        \caption{PRs with behavioral differences.}
        \label{fig:rq1_2_pr_comparison}
    \end{subfigure}

    \vspace{0.8em}

    \begin{subfigure}{\textwidth}
        \centering
        \includegraphics[width=\textwidth]{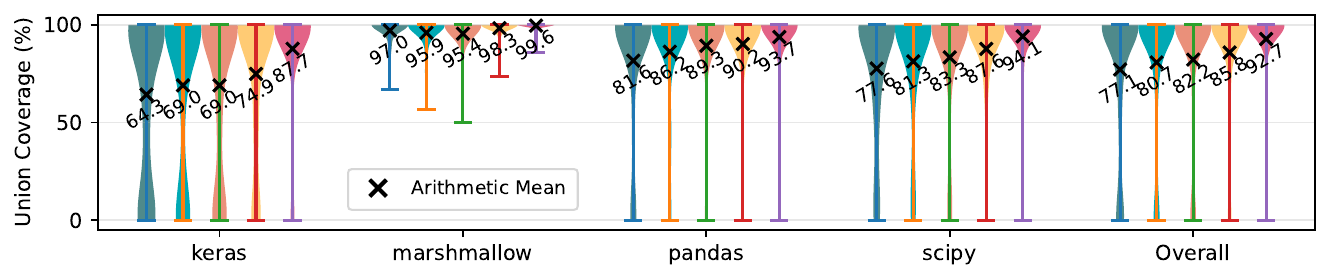}
        \caption{Union coverage.}
        \label{fig:rq1_2_cov_comparison}
    \end{subfigure}

    \caption{Comparison of approaches in terms of PRs with behavioral differences and union coverage.}
    \label{fig:rq1_2_comparison}
\end{figure*}

\subsection{RQ1: Effectiveness}
\label{subsec:rq1}

\subsubsection{Results on Testora Data}
Figure~\ref{fig:rq1_2_comparison} summarizes the results on Testora data across approaches. 
In this research question, we focus on Testora, Testora++, and \name{}.


Figure~\ref{fig:rq1_2_pr_comparison} shows that \name{} identifies the largest number of PRs with behavioral differences, detecting 350 PRs. 
In comparison, Testora and Testora++ detect 251 and 277 PRs, respectively. 
The coverage results in Figure~\ref{fig:rq1_2_cov_comparison} further support this finding. 
\name{} achieves the highest union coverage of 92.7\%, compared with 77.1\% for Testora and 80.7\% for Testora++. 
In addition, Table~\ref{tab:rq1_num_tests} shows that \name{} generates relatively high-quality tests, 
with 27.8\% (4,189/15,055) of the generated tests exposing behavioral differences. 
This proportion is higher than both Testora (15.5\%) and Testora++ (13.6\%), further indicating the effectiveness of \name{} in producing tests that reveal behavioral differences.
Despite generating 4.45$\times$ as many tests as \name{}, Testora++ still achieves lower coverage and detects fewer PRs and tests related to behavioral differences.


\begin{figure}[t]
  \centering
  \includegraphics[width=0.49\textwidth]{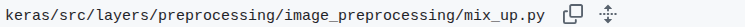}
  \includegraphics[width=0.49\textwidth]{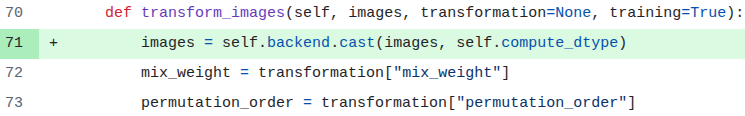}
  \caption{A PR where only \name{} finds different outputs.} 
  \label{fig:rq1_example}
\end{figure} 

\name{} identifies behavioral differences in 70 PRs that are not detected by either Testora or Testora++. 
Figure~\ref{fig:rq1_example} shows one such PR, which introduces a single-line change in function \code{transform\_images}.
Despite this minimal change, both Testora and Testora++ fail to reach the changed line due to missing information on how to invoke the changed function, resulting in \emph{ModuleNotFoundError} in all generated tests.
In contrast, \name{} leverages access information, including an import line and the signature and docstring of the base class, enabling correct invocation and successful coverage of the changed line. 

For the 69 PRs where all three approaches fail to expose behavioral differences,  
the changes can be grouped into several broad categories, which largely correspond to non-functional or low-impact modifications, 
including typing improvements, refactoring, compatibility adjustments, and dtype-related stabilization. 
This suggests that such changes are unlikely to be captured when targeting behavioral changes, which may explain the absence of observable output differences.

\begin{table}[t]
  \centering
  \caption{Number of generated tests across rounds.} 
  \setlength{\tabcolsep}{3.6pt}
  \begin{tabular}{@{}lrrrrrrr@{}}
  \toprule
  \multirow{2}{*}{Approach} & \multicolumn{6}{c}{Generated tests} & \multirow{2}{*}{$\mathit{Num}_\text{test}$} \\
  \cmidrule(lr){2-7}
   & $R_0$ & $R_1$ & $R_2$ & $R_3$ & $R_4$ & Total & \\ 
  \midrule
  Testora & 14,129 & 0 & 0 & 0 & 0 & 14,129 & 2,193\\
  Testora++ & 66,948 & 0 & 0 & 0 & 0 & 66,948 & 9,107\\
  \onlycov{} & 14,120 & 709 & 122 & 28 & 9 & 14,988 & 2,319\\
  \onlycg{} & 14,120 & 0 & 0 & 0 & 0 & 14,120 & 3,525\\
  DiffTestGen & 14,154 & 766 & 101 & 25 & 9 & 15,055 & 4,189\\
  \bottomrule
  \end{tabular}
  \label{tab:rq1_num_tests}
\end{table}

As described in \S\ref{subsec:baselines}, we introduce Testora++ to control for the difference in the number of generated tests. 
Table~\ref{tab:rq1_num_tests} summarizes the number of tests generated by different approaches in each round and the corresponding number of tests that expose behavioral differences. 
The results show that \name{} generates approximately 1,000 more tests than Testora overall, confirming our concern that the comparison could be biased by the larger number of generated tests. 
By increasing the number of generated tests, Testora++ produces 4.45$\times$ as many tests as \name{}, yet it still achieves lower coverage and detects fewer PRs and tests related to behavioral differences.

\subsubsection{Results on ChaCo Data}
Table~\ref{tab:rq1_chaco} summarizes the results of comparing \name{} with ChaCo~\cite{ICSE026_chaco}.
As discussed in \S\ref{subsec:llm}, we also evaluate \name{} using GPT-4o-mini for a fair comparison. 
With GPT-4o-mini, \name{} identifies 21 PRs with behavioral differences, the same number as ChaCo. 
Nevertheless, \name{} achieves a higher union coverage of 64.5\%, outperforming ChaCo's 52.0\%.
Using GPT-5-mini, \name{} further improves its effectiveness, identifying 28 PRs and achieving a union coverage of 76.8\%. 
These results indicate that \name{} matches ChaCo in behavioral difference detection while achieving substantially higher code coverage, 
and that its effectiveness further improves when paired with a more capable LLM.

\begin{table}[t]
  \centering
  \caption{Results on ChaCo data.}
  \setlength{\tabcolsep}{2.4pt}
  \begin{tabular}{@{}lrrrrrr@{}}
    \toprule
    \multirow{3}{*}{Project} & \multicolumn{2}{c}{ChaCo} & \multicolumn{4}{c}{\name{}} \\
    \cmidrule(lr){2-3} \cmidrule(lr){4-7} 
    & \multicolumn{2}{c}{GPT-4o-mini} & \multicolumn{2}{c}{GPT-4o-mini} & \multicolumn{2}{c}{GPT-5-mini}\\
    \cmidrule(lr){2-3} \cmidrule(lr){4-5} \cmidrule(lr){6-7} 
    & $\mathit{Cov}_\text{union}$ & $\mathit{Num}_\text{PR}$  & $\mathit{Cov}_\text{union}$ & $\mathit{Num}_\text{PR}$  & $\mathit{Cov}_\text{union}$ & $\mathit{Num}_\text{PR}$  \\
    \midrule
    pandas  & 48.7\% & 10 & 64.3\% & 11 & 83.4\% & 15 \\
    scipy & 54.9\% & 11  & 64.8\% & 10 & 70.9\% & 13 \\
    \midrule
    Total & 52.0\% & 21 & 64.5\% & 21 & 76.8\% & 28 \\
    \bottomrule
  \end{tabular}
  \label{tab:rq1_chaco}
\end{table}


\subsection{RQ2: Component Contributions}
\label{subsec:rq2}

In this research question, we focus on the results of Testora, \onlycov{}, \onlycg{}, and \name{} shown in Figure~\ref{fig:rq1_2_comparison}. 
These four approaches form an ablation study designed to investigate the contribution of each component of \name{} to its overall effectiveness.


The results show that each added component consistently improves performance, increasing both the number of PRs with behavioral differences and the union coverage. 
In particular, \onlycov{} identifies 261 PRs with behavioral differences and achieves 80.7\% union coverage.
\onlycg{} identifies 330 PRs with behavioral differences and achieves 85.8\% coverage, significantly outperforming Testora (251 PRs and 77.1\% coverage). 
This improvement can be attributed to the inclusion of informative access information in the prompt, which enhances the quality of the generated tests. 
\onlycov{} also yields improvements, yet less pronounced, since it relies on coverage feedback derived from the initial round tests, which are of relatively lower quality. 
The combination of both components in \name{} shows the best results, identifying 350 PRs with behavioral differences and achieving 92.7\% coverage.

\subsection{RQ3: Efficiency}
\label{subsec:rq3}

\begin{table}[t]
  \centering
  \caption{Per-PR cost in terms of tokens, money, and time.}
  \begin{tabular}{@{}llrrr@{}}
    \toprule
    &  & Testora & Testora++ & \name{}\\
    \midrule
    Tokens & Input   & 1,565 & 1,575  & 12,584  \\
    & Output  & 9,052 & 22,540 & 18,677 \\
    & Total  & 10,617 & 24,115 & 31,260 \\
    \midrule
    Cost & Dollars  & 0.018 & 0.045 & 0.041  \\
    \midrule
    Time & Minutes  & 8.92 & 25.36 & 15.86  \\
    \bottomrule
  \end{tabular}
  \label{tab:costs}
\end{table}

Table~\ref{tab:costs} reports the average cost per PR in terms of token usage, monetary cost, and execution time.
Comparing Testora, Testora++, and \name{}, we observe that Testora is the least resource-intensive approach.
This observation, together with the lower $\mathit{Num}_\text{PR}$ detected by Testora and the lower $\mathit{Cov}_\text{union}$ it achieves (see \S\ref{subsec:rq1}), motivates the introduction of Testora++. 
Comparing Testora++ and \name{}, we find that \name{} incurs a lower monetary cost (\$0.041 vs. \$0.045).
This is despite the slightly higher overall token usage (31,260 tokens per PR) than Testora++ (24,115 tokens per PR), because Testora++ uses many more output tokens, which are more expensive than input tokens.
From a time perspective, Testora++ also requires more execution time than \name{}, with an average of 25.36 minutes per PR compared to 15.86 minutes for \name{}. 
In other words, \name{} achieves better effectiveness while being more efficient than Testora++ in terms of both monetary cost and execution time.


\subsection{RQ4: Usefulness}
\label{subsec:rq4}
To assess whether \name{} can help identify regression bugs introduced by PRs, we analyze the 70 PRs for which only \name{} identifies behavioral differences. 
While other PRs reported in~\S\ref{subsec:rq1} may also be regression-related, we focus on these uniquely identified cases to evaluate \name{}'s distinctive regression detection capability.
We adopt the LLM-based classifier from Testora~\cite{pradel2025testora}: Given a test that exposes a behavioral difference and the PR's description, the classifier determines whether the observed difference corresponds to an intended change or a potential regression.

\begin{table}[t]
  \centering
  \caption{The number of PRs, tests, and regressions.}
  \begin{tabular}{@{}lrrrrr@{}}
    \toprule
    & $\mathit{Num}_\text{PR}$ & $\mathit{Num}_\text{test}$ & $\mathit{Num}_\text{test}^\text{classify}$ & $\mathit{Num}_\text{PR}^\text{regression}$\\
    \midrule
    keras       & 24  & 193   & 86   & 4 \\
    marshmallow & 5   & 26    & 13   & 0 \\
    pandas      & 19  & 120   & 55   & 1 \\
    scipy       & 22  & 237   & 94   & 2 \\
    \midrule
    Total  & 70 & 576 & 248 & 7 \\ 
    \bottomrule
  \end{tabular}
  \label{tab:regression}
\end{table}

The first two columns of Table~\ref{tab:regression} present the distribution of the 70 PRs across projects and the corresponding number of tests that expose behavioral differences ($\mathit{Num}_\text{test}$). 
To make the analysis tractable, we randomly sampled up to five tests from each PR for classification. 
This sampling strategy is motivated by two considerations: 
(i) multiple tests associated with the same PR often expose the same behavioral difference, making exhaustive analysis redundant, 
and (ii) classifying all 576 tests and further manual inspection would require substantial effort and time. 
The number of sampled tests, denoted by $\mathit{Num}_\text{test}^{\text{classify}}$, is also listed in Table~\ref{tab:regression}.

The final column in Table~\ref{tab:regression} shows $\mathit{Num}_\text{PR}^\text{regression}$, i.e., the number of PRs that are classified as regression-related.
Among the 70 analyzed PRs, the classifier determines seven to be a regression.
Note that finding most behavioral differences to be intended changes is expected, as developers typically aim to introduce intended changes rather than regressions.
We manually inspect these seven PRs and find that five correspond to actual regression bugs, while the other two are false positives.
Out of the five regressions, two were detected and fixed independently by the developers, i.e., if applied at the right time, \name{} could have helped prevent these regressions from being merged into the codebase.
For the remaining three regressions, we are currently in the process of reporting them to the developers for further investigation.
%

\section{Threats to Validity}
One threat to internal validity is the non-determinism of LLM-based test generation, which may lead to slight variations in the generated tests and exposed behavioral differences, as observed for some individual cases, e.g., \emph{marshmallow} in Figure~\ref{fig:rq1_2_comparison}. 
We mitigate this threat by evaluating on hundreds of PRs, where individual variations are largely averaged out, but repeated runs could still produce slightly different results. 
Another threat concerns our evaluation metrics: behavioral differences are based on observed outputs and runtime errors after filtering flaky tests, and coverage considers executable changed Python lines. 
These metrics are aligned with our goal, but may miss differences that do not manifest in outputs, exceptions, or measured line coverage. 
Similarly, the regression-related results rely on an LLM-based classifier from Testora and manual inspection of sampled cases, which may introduce classification or sampling errors.

A threat to external validity is that \name{} currently focuses on Python code changes in projects with available structure and documentation. 
The results may not directly generalize to other programming languages, projects with sparse documentation, configuration changes, or changes in components implemented outside Python. 
Scalability is another limitation: as code changes grow in size and complexity, extracting informative context, finding useful access paths, and keeping prompts within practical limits become more challenging. 
Finally, our datasets are drawn from prior work and cover a limited set of projects, so effectiveness may vary for projects with different API conventions, testing practices, or dependency environments.

\section{Related Work}
\subsection{Automated Test Generation and Regression Testing}
To exercise program behavior, prior work has explored various directions. 
Much work focuses on improving test coverage~\cite{FSE25_CoverUp, 2026AdvancingCodeCoverage, FSE24_Code_Aware_Prompting, FSE2026_TestGeneralizer, ICSE23_codemosa, ASE2024_HITS}.
For example, TELPA~\cite{2026AdvancingCodeCoverage} improves coverage for hard-to-cover branches. 
While both TELPA and \name{} leverage call graph analysis and prior tests, they differ in both design and objective. 
TELPA constructs call paths for all target methods and retrieves a subset of tests that invoke the entry method, greedily maximizing target-method coverage through incremental coverage gains. 
Unlike this, \name{} combines public API checking and call graph analysis, extracts call paths only for private functions that cannot be invoked directly, and selects a single reference test based on line-distance proximity. 
Furthermore, TELPA aims to improve coverage for hard-to-cover branches, whereas \name{} focuses on exposing behavioral differences introduced by Python code changes.

Other approaches guide LLMs using richer program context, such as test intentions~\cite{ICSE25_IntentionGuided}, 
backward-sliced information~\cite{le2026testweaver} and dynamic symbolic execution~\cite{ISSTA11_eXpress} for test generation, to improve the quality of tests.
TestWeaver~\cite{le2026testweaver} introduces a “close test” notion that leverages control-flow structure to localize tests likely to influence a target line, 
defining closeness via execution of a control-dependent condition and prioritizing the nearest such condition by line distance. 
Unlike this design, \name{} adopts a simpler and more general notion of closeness based purely on line distance, without relying on control-dependence or conditional structures.
%
%

Related efforts also investigate task-specific test generation objectives, including  API testing~\cite{ICSE2025multiagentrestapi}, bug detection~\cite{ISSTA25_CrossProbe, ACL2025_Generation4DetectBugs}, and patch validation~\cite{ICML2025_Otter}. 
To support regression testing under limited time budgets, additional work improves scalability through test case prioritization~\cite{ICSE2002_test_prioritization}, selection~\cite{ICSE2018_Hybrid_selection, AST22_Zhang, issta2025_RTS}, and minimization~\cite{ICSE2018_test_minimization, ICSE2018_Nemo}.
These approaches typically target specific code regions within a single program version (e.g., a function or branch), whereas \name{} focuses on behavioral differences between two program versions. 


\subsection{Test Suite Evolution}
Test suites co-evolve with software systems and require continuous maintenance to remain effective~\cite{FSE2012_TestSuiteEvolution, ICSE2013_TestEvol}. 
Existing work has investigated a range of maintenance activities, including test repair~\cite{TSE2025_TestRepair} and test suite augmentation~\cite{ICSE026_chaco, ICST2022_CodeToTestCoevolution, FSE2024_Meta, OOPSLA2026_BeyondCoverage, ICSE26_Etest}. 
These approaches primarily aim to improve coverage and exercise program behaviors by targeting previously uncovered code lines and integrating newly generated or repaired tests. 
While these approaches focus on overall uncovered code and program behavior, 
\name{} instead focuses specifically on changed code lines and their potential behavioral differences. 

\subsection{Differential Testing and Behavioral Analysis}
Differential testing compares multiple implementations on the same inputs and flags behavioral inconsistencies as potential bugs or semantic divergences, 
making it especially useful when test oracles are unavailable~\cite{mckeeman1998differential, PLDI2011finding}. 
It has been widely applied to compilers. 
Work by Yang et al.~\cite{PLDI2011finding} shows that randomly generated programs executed across multiple C compilers reveal defects through divergent outputs, 
while Lidbury et al.~\cite{lidbury2015many} extend this to OpenCL compilers across heterogeneous architectures.
Beyond compilers, differential testing has been used to validate programming language implementations and analysis tools~\cite{park2021jest, fan2021ast}. 
Unlike JEST~\cite{park2021jest}, which compares multiple JavaScript engines and the ECMAScript specification, and Fan et al.~\cite{fan2021ast}, which compares alternative AST mapping implementations, \name{} focuses on detecting behavioral differences in Python code across versions.
Yang et al.~\cite{ICSE2023_FuzzingAutomaticDifferentiation} apply differential testing across different execution scenarios to detect bugs in deep learning libraries, addressing a different task setting than \name{}. 
In security, differential behavioral analysis detects vulnerabilities by comparing executions under varying inputs, 
including applications of differential fuzzing to deep learning systems~\cite{guo2018dlfuzz} and side-channel analysis~\cite{ICSE2019_DifFuzz}, highlighting subtle input-dependent behavior differences. 
%

\section{Conclusion}
In this work, we presented \name{}, a change-directed LLM-based testing approach for exposing behavioral differences introduced by code changes in Python programs. 
\name{} addresses two key challenges in test generation. 
First, it explicitly focuses on code changes, mitigating the risk of overlooking behavioral differences that are related to code modifications. 
Second, it leverages access information for different categories of changed functions and provides reference tests together with their coverage information to guide LLMs in generating tests, 
addressing the challenge that certain changed code regions are difficult to reach. 
We evaluate \name{} on two datasets, and the results demonstrate its effectiveness in exposing behavioral differences. 
In particular, it identifies behavioral differences in \totalrate{} of PRs while achieving a union coverage of up to \totalunioncov{}. 
Furthermore, the identified behavioral differences can be used to detect regression bugs missed by the best existing approaches. 


\bibliographystyle{IEEEtran}
\bibliography{references}

\end{document}

%% file: tex_files/macros.tex
\newcommand{\totalsize}{463}

\newcommand{\testoracheckedsize}{439}

\newcommand{\chacocheckedsize}{34}

\newcommand{\totalrate}{78.2\%} 
\newcommand{\testoradifferoutputrate}{79.7\%} 
\newcommand{\chacodifferoutputrate}{61.8\%} 

\newcommand{\totaldiffermore}{99}

\newcommand{\totalunioncov}{90.7\%}
\newcommand{\testoraunioncov}{92.7\%}
\newcommand{\chacounioncov}{64.5\%}  

\newcommand{\testoracovmore}{15.6\%} 
\newcommand{\chacocovmore}{12.5\%} 

\newcommand{\onlycov}{Only Coverage Feedback}
\newcommand{\onlycg}{Only Access Information}

%% file: tex_files/changed_function_categorization.tex
\begin{table}[t]
  \centering
  \caption{Access information for each function category.}

  \begin{tabular}{@{}p{6em} p{22em}@{}}
    \toprule
    Category & Access information \\
    \midrule

    \makecell[l]{Changed public\\functions} &
    \vspace{-1.2em}
    \begin{itemize}[leftmargin=*, itemsep=0pt, topsep=0pt]
      \item Import line for changed public function.
      \item Signature and docstring of the class to which the changed public function belongs.
      \vspace{-1.2em}
    \end{itemize}
    \\
    \midrule

    \makecell[l]{Changed private\\functions} &
    \vspace{-1.2em}
    \begin{itemize}[leftmargin=*, itemsep=0pt, topsep=0pt]
      \item \makecell[tl]{Entry function $\mathit{f}_\mathit{entry}$.
      (The publicly accessible\\ function that leads to the changed private function).}
      \item Import line for $\mathit{f}_\mathit{entry}$.
      \item Signature and docstring of $\mathit{f}_\mathit{entry}$.
      \item Signature and docstring of the class to which $\mathit{f}_\mathit{entry}$ belongs.
      \vspace{-1.2em}
    \end{itemize}\\
    \midrule

    \makecell[l]{Changed special\\methods} &
    \vspace{-1.2em}
    \begin{itemize}[leftmargin=*, itemsep=0pt, topsep=0pt]
      \item Import line for the class to which the changed special method belongs. 
      \item Signature and docstring of the class to which the changed special method belongs. 
      \item Usage guideline for the changed special method.
      \vspace{-1.2em}
    \end{itemize} \\
    \midrule
  \end{tabular}

  \label{tab:changed_function_types}
\end{table}

%% file: tex_files/access_info_examples.tex
\begin{figure*}[t]
  \centering
  \small
  \begin{subfigure}[t]{0.98\textwidth}
    \begin{promptMSGNoTitle}{green!15!white}
    To access changed function `style':
    \scode{from pandas.core.frame.DataFrame import style}\\
    * Signature and docstring of class `pandas.core.frame.DataFrame':\\
\scode{class DataFrame(data=None, index: Axes | None=None, columns: Axes | None=None, ...)} \\
    Two-dimensional, size-mutable, potentially heterogeneous tabular data.
Data structure also contains labeled axes ...

    \end{promptMSGNoTitle}
    \caption{Changed public function.}
    \label{fig:access_info_example_public}
  \end{subfigure}

  \vspace{0.5em}
  \begin{subfigure}[t]{0.98\textwidth}
    \begin{promptMSGNoTitle}{cyan!10!white}
      To access changed function \scode{marshmallow.schema.Schema.\_run\_validator}:\\
      Method \#0: by calling function `validate'\\
      To access it: \scode{from marshmallow.schema.Schema import validate}\\
      * Signature and docstring: \\
      \scode{def validate(self, data: (typing.Mapping[str, typing.Any]| ...)) -> dict[str, list[str]]}\\
      Validate `data' against the schema, returning a dictionary of validation errors ...

      * Signature and docstring of class 'marshmallow.schema.Schema':\\
      \scode{class Schema(*, only: types.StrSequenceOrSet | None=None, exclude: types.StrSequenceOrSet=(), ...)} \\
    Base schema class with which to define schemas. 
Example usage: ...\\ - `only': Whitelist of the declared fields to select when
instantiating the Schema... 
\medbreak
      Method \#1: ...
    \end{promptMSGNoTitle}
    \caption{Changed private function.}
    \label{fig:access_info_example_private}
  \end{subfigure}

  \vspace{0.5em}
  \begin{subfigure}[t]{0.98\textwidth}
    \begin{promptMSGNoTitle}{cyan!20!white}
To trigger changed method \scode{marshmallow.types.SchemaValidator.\_\_call\_\_}, which is a special method of the class\\`marshmallow.types.SchemaValidator', access the class with:\\
\scode{from marshmallow.types.SchemaValidator import SchemaValidator}\\
* Hints: ``\_\_call\_\_'' is a special method that makes an object callable, i.e., it allows an instance of a class to be used like a function.\\
* Signature and docstring of class 'marshmallow.types.SchemaValidator': ...
    \end{promptMSGNoTitle}
    \caption{Changed special method.}
    \label{fig:access_info_example_special}
  \end{subfigure}
  \caption{Access information examples for changed public functions, changed private functions, and changed special methods.}
  \label{fig:access_info_examples}
\end{figure*}

%% file: tex_files/call_path_algorithm.tex
\SetKwInput{Input}{Input}
\SetKwInput{Output}{Output}
\begin{algorithm}[t]
  \caption{Extract access information for a $\mathit{f}_{pri}$} 
  \label{alg:access_information}
  \small\Input{$\mathit{func\_info}$: The information of the changed function, specifically its name and location as (file, line, col).\\
  \qquad \quad $k$: The number of expected shortest call paths (top-k).}
  \small\Output{$\mathit{shortests}$: The corresponding access information extracted from the top-k shortest call paths}
  \BlankLine
  $\mathit{all}, \mathit{shortests}, \mathit{access\_info} \gets \emptyset$ \\ 
  $\mathit{checked\_callers}.\mathit{append(func\_info)}$\\
  $\mathit{Queue}.append(\mathit{func\_info}, \mathit{access\_info})$
  \label{alg:init_queue}

  \BlankLine
  \tcp{\footnotesize\textbf{Step 1: Explore call paths to the private function and record access information}}
  \While{Queue}{\label{alg:while_start}
    $\mathit{func\_info}, \mathit{access\_info} \gets \mathit{Queue.extract()}$ \\

    \If{``$\mathit{recursion}$'' in $\mathit{access\_info}$}{ \label{alg:terminate_loop_starts}
      \textbf{continue} \quad \tcp{\footnotesize{discard the path}} 
    } \label{alg:terminate_loop_ends}

    $\mathit{f}_\mathit{entry}^{tmp} = \mathit{all}_\mathit{call\_path}[0] $ \\
    \If{$\mathit{isPubliclyAccessible}(\mathit{f}_\mathit{entry}^{tmp})$}{\label{alg:ends_check_starts}
      $\mathit{all}.\mathit{append(access\_info)}$\\
      \textbf{continue} \label{alg:ends_check_ends}
    } 

    $\mathit{callers} \gets \mathit{getFunctionCallers}(\mathit{func\_info})$ \label{alg:get_caller}\\
    \For{$\mathit{caller}$ in $\mathit{callers}$}{ 
      $\mathit{caller\_info}, \mathit{access\_info} \gets \mathit{checkCaller}(\mathit{caller})$ \\
      \If{$\mathit{caller}$ in $\mathit{checked\_callers}$}{ \label{alg:loop_handle_starts}
          $\mathit{access\_info}^{'} \gets \mathit{updateAccessInfo}(``\mathit{recursion}$'') \\
          \textbf{continue} \label{alg:loop_handle_ends} 
      }
      $\mathit{checked\_callers}.\mathit{append(caller)}$
      $\mathit{access\_info}^{'} \gets \mathit{updateAccessInfo}(\mathit{access\_info})$ \\
      $\mathit{Queue.append}(\mathit{caller\_info}, \mathit{access\_info}^{'})$ \label{alg:update} 
    }
  }\label{alg:while_ends}

  \BlankLine
  \tcp{\footnotesize\textbf{Step 2: Select the shortest call path and its corresponding access information}} 
  \tcp{\footnotesize 2.1 Select the shortest for each $\mathit{f}_\mathit{entry}$}
  $\mathit{Shortests}_{entry\_level} \gets \emptyset$ \label{alg:2.1_starts} \\ 
  $\mathit{entry\_to\_paths} \gets groupByEntryFuncs(\mathit{all})$\\ 
  
  \For{$\mathit{ep}$ in $\mathit{entry\_to\_paths}$}{ 
    $\mathit{Shortests}_{entry\_level}.append(min(ep))$\\
  } \label{alg:2.1_ends}

  \tcp{\footnotesize 2.2 Select the top-k shortest from 2.1}
  $Shortests \gets getShortestK(\mathit{Shortests}_{entry\_level}, k)$ \label{alg:2.2}\\
  
  \Return{$Shortests$}
\end{algorithm}

%% file: tex_files/prompt_info.tex
\begin{table}[t]
  \caption{Initial and coverage feedback prompt information.}
  \label{tab:prompt_info}

  \begin{tabular}{@{}l p{15em} p{3em} p{8em}@{}}
    \toprule
    & Information & Initial prompt & Coverage feedback prompt \\
    \midrule
    1 & Name of the project  & \ding{51} & \ding{51} \\
    \midrule
    2 &  Diff of the code change   & \ding{51} &  \ding{51}\\
    \midrule
    3 &  Fully qualified names of the changed functions   & \ding{51}  & \ding{51} \\
    \midrule
    4 &  Bodies of non-test functions in old and new versions  & \ding{51}*  &  \ding{55}\\
    \midrule
    5 &  Reference test  & \ding{55} &  \ding{51} \\
    \midrule
    6 &  Commented non-test function bodies   & \ding{55} &  \ding{51}  $^\#$ \\
    \midrule
    7 &  Access information  & \ding{51}* &  \ding{51} * $^\#$ \\
    \midrule
    8 & Bodies of test functions in old and new versions  & \ding{51}* & \ding{51} * $^\#$ \\
    \bottomrule
  \end{tabular}
  \footnotesize

  \vspace{.2cm}
* Included only if the prompt remains within the target length limit.\\
$^\#$ Only information associated with uncovered changed functions.
\end{table}

%% file: tex_files/refer_test_example.tex
\begin{figure}[t]
  \centering
  \small
  \begin{promptMSGNoTitle}{violet!10!white} 
Reference test: 
\begin{codeblock}
import tensorflow as tf
from keras.src.backend.tensorflow.numpy import eye
\end{codeblock}
\medbreak
\begin{codeblock}
N, M, k = 5, 4, -1
x = eye(N, M=M, k=k, dtype=tf.float32)
print("Example: eye(\{\}, \{\}, k=\{\})".format(N, M, k))
print("shape:", x.shape)
print(x.numpy())
\end{codeblock}
\medbreak
The following are the old and the new versions of the uncovered affected functions:\\
Old version:\\
The following are the affected functions from file\\ ``keras/src/backend/tensorflow/numpy.py'':

\begin{codeblock}
def eye(N, M=None, k=0, dtype=None):
    dtype = dtype or config.floatx()
    if not M: \textcolor{teal}{\textcolor{teal}{# COVERED}}
        M = N \textcolor{violet}{\textcolor{violet}{# TO_COVER}}
    # Making sure N, M and k are `int`
    N, M, k = int(N), int(M), int(k) \textcolor{teal}{# COVERED}
    if k >= M or -k >= N: \textcolor{teal}{# COVERED}
        return zeros([N, M],dtype=dtype)\textcolor{violet}{# TO_COVER}
    ...
\end{codeblock}

New version: ...

\end{promptMSGNoTitle}

  \caption{A reference test and its coverage information.}
  \label{fig:reference_test}
\end{figure}